\documentclass[aps,pra,twocolumn]{revtex4-1}

\usepackage[utf8]{inputenc}
\usepackage{xparse}
\usepackage{mathtools}
\usepackage{physics}
\usepackage{hyperref}
\usepackage[capitalise]{cleveref}
\usepackage{subdepth}
\usepackage{glossaries}

\AtBeginDocument{
  \usepackage{booktabs}
}

\crefalias{subequation}{equation}
\crefname{section}{Sec.}{Section}
\crefalias{subsection}{section}

\DeclareMathOperator\PTop{\mathcal{PT}}

\DeclareMathOperator\Hamiltonian{\mathcal{H}}

\newcommand{\imag}{\mathrm{i}\mkern1mu} 
\newcommand{\PT}{{\texorpdfstring{\ensuremath{\PTop}}{PT}}}
\newcommand{\jR}{j_\mathrm{R}}
\newcommand{\jL}{j_\mathrm{L}}
\newcommand{\kS}{k_\mathrm{S}}
\newcommand{\psiI}{\psi_\mathrm{I}}
\newcommand{\psiR}{\psi_\mathrm{R}}

\newcommand*\subrefformat[2]{#1\,(#2)}

\NewDocumentCommand\subref{smm}{%
  \subrefformat{\IfBooleanTF{#1}{\ref*{#2}}{\ref{#2}}}{#3}%
}

\NewDocumentCommand\csubref{smm}{%
  \subrefformat{\IfBooleanTF{#1}{\cref*{#2}}{\cref{#2}}}{#3}%
}

\NewDocumentCommand\csubrefs{smm}{%
  \subrefformat{\IfBooleanTF{#1}{\namecrefs*{#2} \labelcref{#2}}%
    {\namecrefs{#2} \labelcref{#2}}}{#3}%
}

\NewDocumentCommand\Csubref{smm}{%
  \subrefformat{\IfBooleanTF{#1}{\Cref*{#2}}{\Cref{#2}}}{#3}%
}

\NewDocumentCommand\Csubrefs{smm}{%
  \subrefformat{\IfBooleanTF{#1}{\nameCrefs*{#2} \labelcref{#2}}%
    {\nameCrefs{#2} \labelcref{#2}}}{#3}%
}

\NewDocumentCommand\labelcsubref{smm}{%
  \subrefformat{\IfBooleanTF{#1}{\labelcref*{#2}}{\labelcref{#2}}}{#3}%
}

\newcommand\ac[1]{\gls*{#1}}

\newcommand\acrodef[3][\empty]{%
  \ifx{#1}\empty%
    \newacronym{#2}{#2}{#3}%
  \else%
    \newacronym[#1]{#2}{#2}{#3}%
  \fi%
}

\acrodef{TMS}{two-mode system}

\begin{document}


\begin{abstract}

R. Labouvie et al.\ [Phys.\ Rev.\ Lett.\ 116, 235302 (2016)] have described an experiment with a weakly interacting Bose-Einstein condensate trapped in a one-dimensional optical lattice with localized loss created by a focused electron beam. We show that by setting suitable initial currents between neighboring sites it is possible to create $\mathcal{PT}$-symmetric quasi-stationary and $\mathcal{PT}$-symmetry broken decaying states in an embedded two-mode subsystem. This subsystem exhibits gain provided by the coupling to the reservoir sites and localized loss due to the electron beam, and shows the same dynamics as a non-Hermitian two-mode system with symmetric real and antisymmetric imaginary time-independent potentials, except for a proportionality factor in the chemical potential. We also show that there are three other equivalent scenarios, and that the presence of a localized loss term significantly reduces the size of the condensate required for the realization.
\end{abstract}

\title{Realization of $\mathcal{PT}$-symmetric and $\mathcal{PT}$-symmetry broken states in static optical lattice potentials}
\author{Felix Kogel}
\author{Sebastian Kotzur}
\author{Daniel Dizdarevic}
\email{daniel.dizdarevic@itp1.uni-stuttgart.de}
\author{J\"org Main}
\author{G\"unter Wunner}
\affiliation{Institut f\"ur Theoretische Physik 1, Universit\"at Stuttgart,
             70550 Stuttgart, Germany}
\date{\today}

\maketitle

\section{Introduction}

In 1998 Bender and Boettcher \cite{Bender1998} introduced a new class of Hamiltonians $\Hamiltonian$, which are invariant with respect to the combined effect of the parity operator $\mathcal{P}$ and the time-reversal operator $\mathcal{T}$, that is $\commutator{\Hamiltonian}{\PT} = 0$, without being necessarily invariant with respect to either of them. This allows for non-Hermitian Hamiltonians with entirely real spectra within a so-called {\PT}-symmetric regime, and otherwise an eigenvalue structure with complex conjugate pairs called {\PT}-symmetry broken.

There is a wide range of applications for {\PT}-symmetric systems, as non-Hermitian Hamiltonians are particularly suited to effectively describe open quantum systems (e.g.\ see \cite{Graefe2008}). They can, for example, be used to describe delocalization transitions in condensed matter \cite{Hatano1996}, or for the investigation of population biology \cite{Nelson1998} and exceptional points \cite{Gutoehrlein2016,Schnabel2017,Pan2019,Sweeney2019}. Furthermore, the concept of {\PT} symmetry can be applied to the fields of laser modes \cite{Chong2011, Ge2011, Liertzer2012}, electronic circuits \cite{Schindler2012, Schindler2011, Ramezani2012}, microwave cavities \cite{Bittner2012}, or for the realization of unidirectional invisibility \cite{Lin2011, Loran2016, Lv2017}.

A direct observation of {\PT} symmetry is possible in optical systems \cite{Guo2009, Rueter2010} due to the mathematical equivalence of the wave equation of electrodynamics in paraxial approximation and the Schr\"odinger equation. By considering light propagation in two wave guides the transition between the {\PT}-symmetric and the {\PT}-symmetry broken regime can be investigated. However, up to date an experimental observation of {\PT} symmetry in a quantum mechanical system is still lacking.

A promising experimental procedure for the realization of {\PT}-symmetric quantum systems was proposed by considering a Bose-Einstein condensate in an optical double-well potential \cite{Klaiman2008}. As shown in Refs.~\cite{Cartarius2012,Dast2013a,Dast2013b} the formalism of {\PT} symmetry can be applied to such a nonlinear quantum system, which can develop stable {\PT}-symmetric states \cite{Haag2014}. In more recent works the use of bounded and unbounded states \cite{Single2014} or coupling to another Bose-Einstein condensate \cite{Gutoehrlein2015} were suggested to provide a coherent in- and out-coupling of particles. However, both methods are difficult to realize experimentally.

Kreibich et al.\ \cite{Kreibich2013,Kreibich2014,Kreibich2016} proposed an experiment based on time-dependent optical lattices \cite{Henderson2009}, in which the wells are loaded with Bose-Einstein condensates \cite{Peil2003}. In this way a two-mode system embedded into a larger multi-well system is created, which shows {\PT}-symmetric dynamics in the mean-field approximation and beyond \cite{Dizdarevic2018,Mathea2018}. Although this approach effectively allows for the realization of {\PT} symmetry, the experimental setup is quite demanding and currently hardly realizable due to the time-dependent optical potentials.

In the present paper we will focus on an experiment with a time-independent optical lattice and localized Bose-Einstein condensates of $^{87}\text{Rb}$ atoms \cite{Labouvie2016}. An electron beam can be used to create local losses at specific lattice sites \cite{Barontini2013}. In the following we will propose a modification of this experiment to realize {\PT}-symmetric and {\PT}-symmetry broken states in the mean-field approximation. It is sufficient to describe the system in the mean field, as effects beyond the mean-field theory do not play a role due to the large number of atoms present in the experiment.

\section{Theory}
\label{sec:theory}

\subsection{Two-mode system}

To demonstrate the characteristics of {\PT} symmetry a two-well system filled with ultracold Bose-Einstein condensates is considered where the imaginary part of a complex potential describes the in- and out-coupling of particles. If the potential barrier separating the two wells is high enough, the respective wave functions can be assumed to be localized, so that the system is discrete. By using dimensionless units ($\hbar = m = 1$), the theoretical description of localized Bose-Einstein condensates in the {\PT}-symmetric double well is given by the discrete Gross-Pitaevskii equation (e.g.\ see \cite{Graefe2012})
\begin{align}
  \imag \pdv{t}
  \mqty(\psi_1 \\ \psi_2)
  = \mqty(
    g \abs{\psi_1}^2+\imag\gamma & -J \\
    -J & g \abs{\psi_2}^2-\imag\gamma
  ) \mqty(\psi_1 \\ \psi_2) ,
  \label{eq:GPE_2}
\end{align}
to which we will refer as the \ac{TMS} in the following. While the real part of the potential is symmetric, the imaginary part is anti-symmetric. The coherent coupling with the environment is given by the gain and loss factor $\gamma$, which effectively represents a complex potential. The factor $J$ is the coupling and describes the tunneling of the particles between the two wells. The Hamiltonian \labelcref{eq:GPE_2} represents a nonlinear system with the corresponding strength of nonlinearity $g = 4\pi a N$, where $N$ is the particle number and $a$ describes the scattering length according to the s-wave scattering in Bose-Einstein condensates \cite{Pethick2008}.

With the use of mean-field wave functions $\psi_i = \sqrt{n_i}\exp(\imag \varphi_i)$, where $n_i = \abs{\psi_i}^2$ is the number of particles and $\varphi_i$ the phase of the condensate in the corresponding lattice site, the system in \cref{eq:GPE_2} can be solved. Under the assumption of a symmetric occupation distribution with $n_i = n_0$ the solution of the {\PT}-symmetric system is given by
\begin{equation}
  \vb{\phi} = \mqty(
    \sqrt{n_0}\exp(\imag\varphi) \\ \sqrt{n_0} \exp(-\imag\varphi)
  )
  \label{eq:2M_SS}
\end{equation}
with the phase
\begin{equation}
	\varphi = -\frac{1}{2} \arcsin\qty(\frac{\gamma}{J}).
  \label{eq:varphi_theo}
\end{equation}
The chemical potentials
\begin{equation}
  \mu = gn_0 \pm \sqrt{J^2-\gamma^2}
  \label{eq:mu_theo}
\end{equation}
of these eigenstates are purely real for $\abs{\gamma} \leq J$. In this case the time-evolved solutions $\psi_i(t) = \phi_i \exp(-\imag \mu t)$ are stationary and constitute the so-called {\PT}-symmetric solutions. For larger values of $\gamma$ the eigenvalues contain an imaginary part resulting in a time-dependent norm of the states, so that the {\PT} symmetry is broken.

In the following investigations our aim is to realize the stationary symmetric solutions of the system. Therefore, the accessible observables
\begin{subequations}\label{eq:def:c_kl-j_kl}
	\begin{align}
	c_{k,l} &= 2 \sqrt{n_kn_l} \cos(\varphi_l - \varphi_k), \label{eq:def:c_kl}\\
	j_{k,l} &= 2 J \sqrt{n_kn_l} \sin(\varphi_l - \varphi_k), \label{eq:def:j_kl}
	\end{align}
\end{subequations}
i.e.\ the correlation $c$ and the net current $j$, are introduced in order to describe important properties of the dynamics of the {\PT}-symmetric states. The corresponding characteristic values of the \ac{TMS}
\begin{subequations}\label{eq:TMS:c-j}
	\begin{align}
	c_{1,2} &=  2 n_0 \sqrt{1 - \qty(\frac{\gamma}{J} )^2},
	\label{eq:PTSS:a} \\
	j_{1,2} &= 2n_0 \gamma,
	\label{eq:PTSS:b}
	\end{align}
\end{subequations}
are time-independent and depend only on the phase difference of the two components of the wave function \labelcref{eq:2M_SS}.

\subsection{Complex-extended wave functions} \label{subsec:Compl-ext-wave-funcs}
Beyond the {\PT}-symmetric regime with $\abs{\gamma} > J$, the resulting complex eigenvalues \labelcref{eq:mu_theo} and time-dependent norm cause a time dependence of the Hamiltonian in \cref{eq:GPE_2}. Nevertheless, for a time-independent Hamiltonian with a vanishing nonlinearity $g=0$, the wave functions \labelcref{eq:2M_SS} with a complex phase present solutions of the Schrödinger equation \labelcref{eq:GPE_2}. By using the general relation
\begin{align} \label{eq:complex_arcsin}
  \arcsin(\alpha) &= -\imag \ln\qty( \imag\alpha \pm \sqrt{1-\alpha^2} ) \notag\\
  	&= \frac{\pi}{2} - \imag \ln(\alpha \pm \sqrt{\alpha^2-1})
\end{align}
for $\abs{\alpha} > 1$, the phase \labelcref{eq:varphi_theo} turns into
\begin{align}
	\varphi = -\frac{\pi}{4}
    + \frac{\imag}{2} \ln{\qty(\alpha \pm \sqrt{\alpha^2-1})}
\end{align}
and has the effect of shifting the initial occupations of each well so that the wave function reads
\begin{equation}
  \vb{\phi} = \mqty(
    \sqrt{n_0\qty(\alpha \pm \sqrt{\alpha^2-1} )} \mathrm{e}^{-\imag\pi/4} \\
    \sqrt{n_0\qty(\alpha \mp \sqrt{\alpha^2-1} )} \mathrm{e}^{\imag\pi/4}
  ) ,
  \label{eq:phi_PT-broken}
\end{equation}
where $\alpha = \gamma / J$. The two possible solutions in \cref{eq:phi_PT-broken} correspond to the two solutions \labelcref{eq:mu_theo}. For every such solution there are either exponentially increasing, or exponentially decreasing particle numbers
\begin{equation} \label{eq:n_i(t)_theo-broken}
  n_i(t) = n_i(0) \mathrm{e}^{2 \Im(\mu) t} ,
\end{equation}
which correspond to the {\PT}-symmetry broken states.

\subsection{Open few-mode model}

An experimental realization of the {\PT}-symmetric and {\PT}-symmetry broken states of the \ac{TMS} requires a coherent in- and out-coupling of particles. It was shown that an out-coupling can be easily obtained with a focused electron beam \cite{Barontini2013}. An injection of particles can be realized by embedding the \ac{TMS} into an optical lattice filled with Bose-Einstein condensates, which acts as a particle reservoir and allows for a steady current of particles into the system.

The mean-field description of a general open few-mode model is again given by the Gross-Pitaevskii equation,
\begin{align}
  \imag \pdv{t} \psi_k
    &= -J \psi_{k-1} - J \psi_{k+1} + g \abs{\psi_k}^2\psi_k \nonumber \\
    &\quad + \mu_k \psi_k - \imag \frac{\gamma_k}{2} \psi_k ,
  \label{eq:open_few-mode_model}
\end{align}
where $\psi_k$ represents a Bose-Einstein condensate localized in the lattice site $k$ of a one-dimensional optical lattice with the corresponding onsite energy $\mu_k$. The parameters $\gamma_k$ describe local in- or out-couplings of particles depending on their signs. The strength of the interaction $g$ is assumed to be equal to the strength of the nonlinearity $g$ of the {\PT}-symmetric \ac{TMS} at each lattice site. In order to ensure a simple realization, the tunneling coupling $J$ between neighboring lattice sites is assumed to be equal at all sites.

The differential equations \labelcref{eq:open_few-mode_model} describe the experimental setup in \cite{Labouvie2016}. In this experiment, roughly $45\,000$ $^{87}\text{Rb}$ atoms were placed in a one-dimensional optical lattice, so that around $700$ particles are located at each lattice site in the center of the potential. It was shown that the use of a strong loss term leads to the creation of a stationary state in the corresponding lattice site due to the quantum Zeno effect \cite{Misra1977, Kofman2000}. However these stationary states do not have the characteristics of the solutions in the {\PT}-symmetric \ac{TMS} in equation \labelcref{eq:TMS:c-j}. In the following, it is shown that the setting of appropriate initial phases and particle numbers of the condensates leads to stationary and exponentially decaying states with the {\PT}-symmetric and {\PT}-symmetry broken characteristics of the \ac{TMS}, respectively.

\subsubsection{{\PT}-symmetric regime}
\label{subsubsec:PT-symm-regime}

\begin{figure}
	\centering
  \includegraphics[width=\columnwidth]{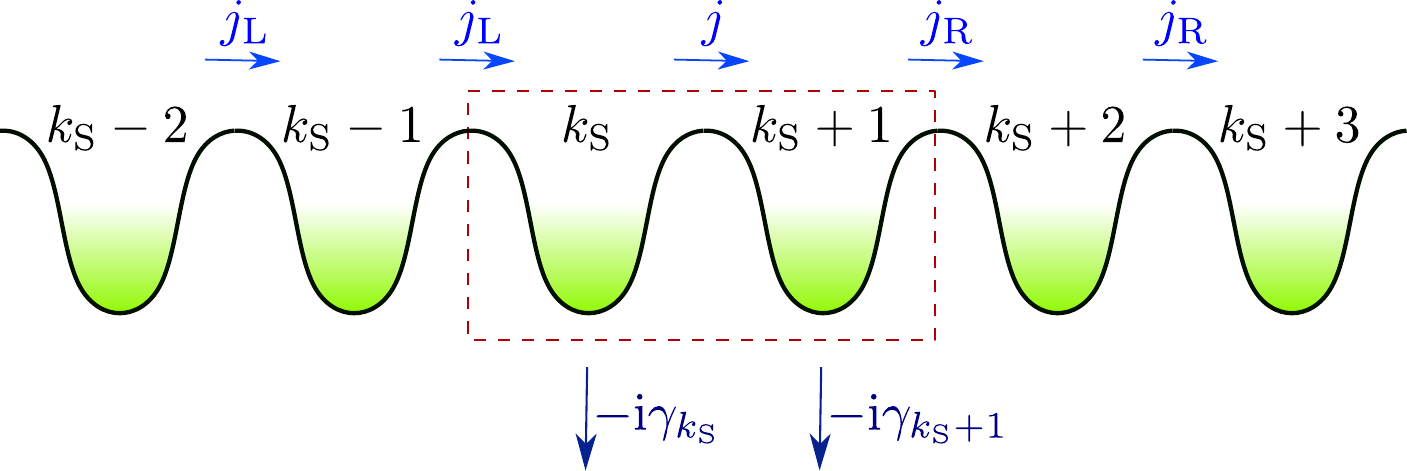}
	\caption{The inner lattice sites with the currents $\jL,j$ and $\jR$ between them and two local loss terms $-\imag \gamma_{\kS}$ and $-\imag \gamma_{\kS+1}$ in the subsystem which is bounded by the dashed rectangle.}
	\label{fig:fig1}
\end{figure}

In the following investigations the goal is to obtain the {\PT}-symmetric states in the two sites of the subsystem labeled with $\kS$ and $\kS+1$. In particular the physically observable parameters, that is the particle number and the current, are desired to be realized as the constant characteristic values of the \ac{TMS}. As the onsite energy leads to a shift in the energy of the system and has no influence on the dynamics it is set to zero. The characteristics of the subsystem have to fulfill the conditions of the \ac{TMS}, viz.
\begin{subequations}
  \label{eq:initial_parameters_symm_solution}
	\begin{align}
	  n_{\kS} = n_{\kS+1}
      &\overset{!}{=} n_0 , \\
	  j_{\kS,\kS+1}
      &\overset{!}{=} j = 2n_0 \gamma .
	\end{align}
\end{subequations}

\begin{table}[t]
  \centering
  \caption{The different possibilities for the values of the currents $\jL$ and $\jR$ with the local gain and loss terms $\gamma_{\kS}$ and $\gamma_{\kS+1}$ assuming that the current $j$ in the subsystem is positive, $j>0$. Positive values of $\gamma_{\kS}$ and $\gamma_{\kS+1}$ correspond to the loss of particles, while negative values correspond to particle gain.}
  \begin{tabular}{cccccccc}
    \toprule
    &&&&& \multicolumn{3}{c}{orientations of} \\
    $\jL$ & $\jR$ & $n_0\gamma_{\kS}$ & $n_0\gamma_{\kS+1}$
      && $\jL$ & $j$ & $\jR$ \\
    \cmidrule{1-4} \cmidrule{6-8}
    $+j$  & $-j$ &  0 & $2j$
      && $\longrightarrow$ & $\longrightarrow$ & $\longleftarrow$ \\
    $+j$  & $+j$ &  0 & 0
      && $\longrightarrow$ & $\longrightarrow$ & $\longrightarrow$ \\
    $-j$  & $+j$ &  $-2j$ & 0
      && $\longleftarrow$ & $\longrightarrow$ & $\longrightarrow$ \\
    $-j$  & $-j$ &  $-2j$ & $2j$
      && $\longleftarrow$ & $\longrightarrow$ & $\longleftarrow$ \\
    \bottomrule
  \end{tabular}
  \label{tab:current_possibilities_jL_j_jR}
\end{table}

The other currents $j_{k,k+1}$ to the left of the subsystem are considered as equal and will be called $\jL$. The same holds for the currents to the right, which will consequently be called $\jR$. This special experimental setup with constant localized loss terms $-\imag \gamma_{\kS}$ and $-\imag \gamma_{\kS+1}$, which only exist in the subsystem, is shown in \cref{fig:fig1}. If these parameters $\gamma_k$ are negative, they act as a gain of particles in these lattice sites instead. In the {\PT}-symmetric \ac{TMS} the occupation in each well is a time-independent constant, as there is the same amount of particles that are coupled in and out. For this reason it has to be ensured that the occupations are constant. Since
\begin{equation}
	\pdv{t} n_k = j_{k-1,k} - j_{k,k+1} - \gamma_k n_k ,
  \label{eq:deriv_n_k_general}
\end{equation}
which follows from \cref{eq:open_few-mode_model}, the time-independence of the particle numbers of the subsystem yield the parameters
\begin{subequations}\label{eq:gamma_k_S_gamma_k_Sp1_theo}
	\begin{align}
	 \gamma_{\kS}	&= \frac{\jL -j}{n_0} , \\
	 \gamma_{\kS+1}	&= \frac{j - \jR }{n_0} .
	\end{align}
\end{subequations}
To fulfill the requirements of time-independent currents and correlations in all lattice sites it can be shown by using \cref{eq:open_few-mode_model} that all currents must have the same absolute values
\begin{equation}
  \abs{\jL} \overset{!}{=}	j \overset{!}{=} \abs{\jR} ,
\end{equation}
assuming there is a positive current $j$ in the subsystem, i.e.\ from left to right (cf.\ \cref{fig:fig1}). Moreover, all initial occupations have to be equally distributed throughout the lattice with $n_k(t=0) = n_0$ to produce the desired dynamics. As a result, there exist four different possibilities for the orientations of the currents $\jL$ and $\jR$ as shown in \cref{tab:current_possibilities_jL_j_jR}. According to \cref{eq:def:j_kl}, the derived initial values for the currents and particle numbers yield the phase differences
\begin{equation}
\varphi_{k+1} - \varphi_{k}
= \arcsin\qty(\frac{j_{k,k+1}}{2J \sqrt{n_k n_{k+1}}}) ,
\label{eq:ansatz_diff_phi_k}
\end{equation}
which have to be prepared initially. Consequently, all initial phases in \cref{eq:ansatz_diff_phi_k} have to exhibit the same phase differences $\pm 2\varphi$ of the \ac{TMS} in \cref{eq:varphi_theo} to ensure the stationary dynamics.

\subsubsection{{\PT}-symmetry broken regime}
\label{subsubsec:PT-broken-regime}

In a manner similar to \cref{subsec:Compl-ext-wave-funcs}, we apply the approach of using complex phases to realize the {\PT}-symmetry broken states of the \ac{TMS}. It is expected that the characteristic dynamics for $\gamma > J$ can also be created by a preparation of the initial phases and occupations. To obtain analytic solutions in the following, the nonlinearity is set to zero, i.e.\ $g = 0$. Since $g$ is freely adjustable by Feshbach resonances \cite{Inouye1998}, this is no restriction with respect to experimental realizability.

The ansatz \labelcref{eq:ansatz_diff_phi_k} with the current $j$ of the {\PT}-symmetric states in \cref{eq:PTSS:b} yields the phase differences
\begin{equation}
  \varphi_{k+1} - \varphi_{k}
  = \begin{cases}
    \arcsin\qty( s_1 \dfrac{\gamma}{J} )	& \text{for } k < \kS , \\
    \arcsin\qty( \dfrac{\gamma}{J} )	& \text{for } k = \kS , \\
    \arcsin\qty( s_2 \dfrac{\gamma}{J} )	& \text{for } k > \kS ,
  \end{cases}
  \label{eq:phi_left_right_broken}
\end{equation}
where the signs $s_1$ and $s_2$ can be selected independently, which corresponds to the four possibilities for the currents shown in \cref{tab:current_possibilities_jL_j_jR}. Since $\gamma > J$, the initial complex phases have an impact on the initial occupations (see \cref{eq:complex_arcsin}), so that the time-dependent particle numbers of the subsystem follow as
\begin{subequations}\label{eq:ansatz1_n_ks_n_ksp1}
	\begin{align}
  	n_{\kS}(t) &= n_0 \qty(
        \alpha \pm \sqrt{\alpha^2-1}
      ) \mathrm{e}^{2\Im(\mu)t} ,
  	\label{eq:ansatz1_n_ks} \\
  	n_{\kS+1}(t) &= n_0 \qty(
        \alpha \mp \sqrt{\alpha^2-1}
      ) \mathrm{e}^{2\Im(\mu)t}
  	\label{eq:ansatz1_n_ksp1}
	\end{align}
\end{subequations}
with $\mu = \pm\sqrt{J^2-\gamma^2}$.
If this is transferred to all lattice sites, all neighboring occupations obey the relation
\begin{equation}\label{eq:exp-distr-particle-num}
  n_k(t) = n_{k+1}(t) \qty( \alpha \pm \sqrt{\alpha^2-1} )^2 .
\end{equation}
Thus, the effective initial preparation consists of exponentially distributed particle numbers and the same absolute phase differences of $\pi/2$ between all lattice sites.

\section{Results}

\subsection{Realizing {\PT} symmetry}

Using the results of \cref{sec:theory} a one-dimensional lattice with $50$ sites is considered. The subsystem $\qty{\kS, \kS+1}$ with $\kS = 25$ is located in the middle of this lattice with uniformly distributed optical characteristics. Without loss of generality we choose units such that $J=1$. Further, the nonlinearity $g$ is now set to zero both in the {\PT}-symmetric and the {\PT}-symmetry broken regimes. As the mean-field dynamics is independent of the particle number, the initial occupations can be set to $n_k(t=0) = 0.5$. According to the four possibilities for the currents and their respective gain and loss terms $\gamma_{\kS}$ and $\gamma_{\kS+1}$ in \cref{tab:current_possibilities_jL_j_jR}, the phase differences must be equal to the values given in \cref{eq:ansatz_diff_phi_k}. Since the absolute phase shift is arbitrary, the phases of the subsystem are set to the values of the \ac{TMS} given in \cref{eq:varphi_theo}, $\varphi_{\kS} = -\varphi_{\kS+1} = \varphi$.

\begin{figure}
	\centering
	\includegraphics[width=\columnwidth]{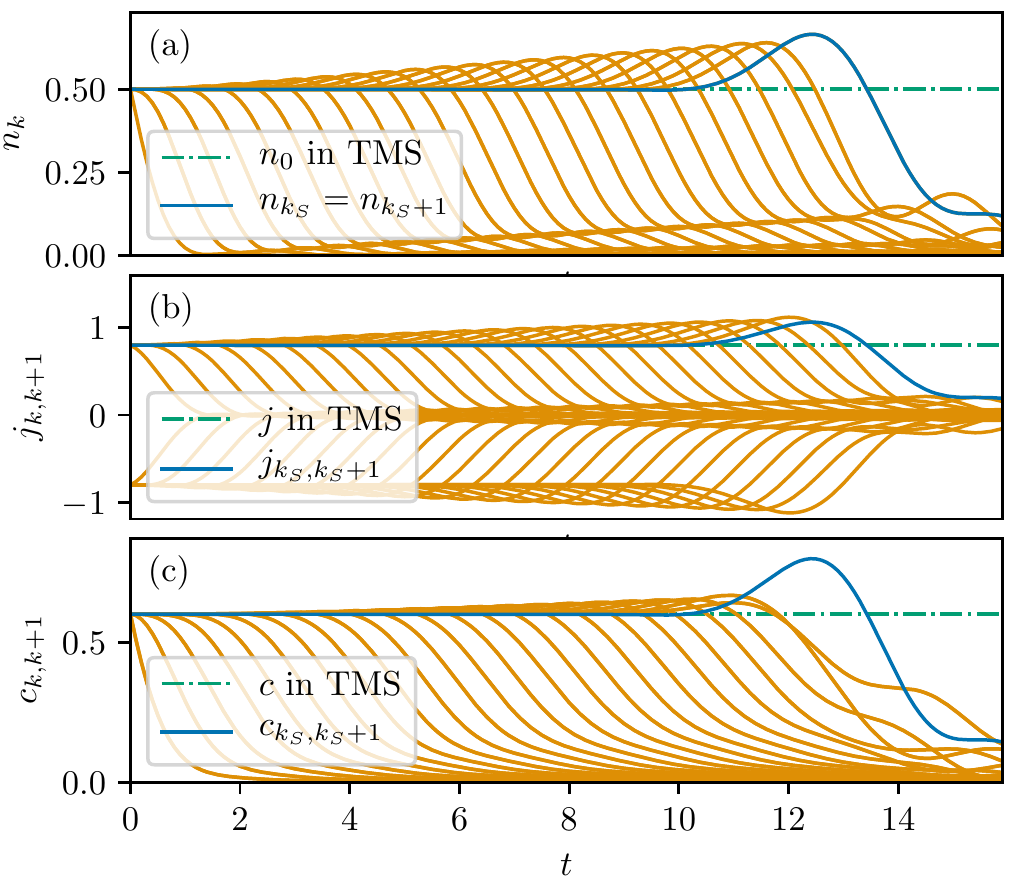}
	\caption{The dynamics of the occupations, the currents, and the correlations of all 50 lattice sites for the initial currents $\jL = j = -\jR$ with the gain and loss term $\gamma = 0.8$ and nonlinearity $g = 0$. The occupations of the subsystem start to deviate significantly from the initial occupations  at $t \approx 10$. The curves which decrease at a smaller time scale belong to the outer sites.}
	\label{fig:fig2}
\end{figure}

To begin with, the first case for the currents in \cref{tab:current_possibilities_jL_j_jR} is considered: The directions of all currents are aligned towards the subsystem in which, only in the right-hand site, a loss term with $\gamma_{\kS+1} = 4 \gamma$ exists. The resulting dynamics of the occupations and the correlations as well as the currents of all lattice sites is shown in \cref{fig:fig2} for the gain and loss factor $\gamma = 0.8$.

It is noticeable that the occupation of the inner lattice sites in \csubref{fig:fig2}{a} remains approximately constant while the outer sites decrease one after another. Furthermore, all curves develop a local maximum. This is due to the fact that the tunneling currents exceed the initially set currents. The constant particle numbers of the subsystem remain at the constant value $n_0$ of the \ac{TMS} until $t\approx 8$ when the adjoining sites can no longer maintain the appropriate current and thus the {\PT} symmetry breaks down.

The currents and correlations in the \csubrefs{fig:fig2}{b} and \labelcsubref{fig:fig2}{c} exhibit a similar behavior because they are directly correlated to the occupations. With the gain and loss factor that is used in \cref{fig:fig2} the characteristic values of the {\PT}-symmetric regime are given by $c_{\kS,\kS+1} = 0.6$ and $j_{\kS,\kS+1} = 0.8$ according to \cref{eq:PTSS:a,eq:PTSS:b}. These theoretical expectations of the \ac{TMS} are depicted by the dash-dotted straight lines.

\begin{figure}
  \centering
  \includegraphics[width=\columnwidth]{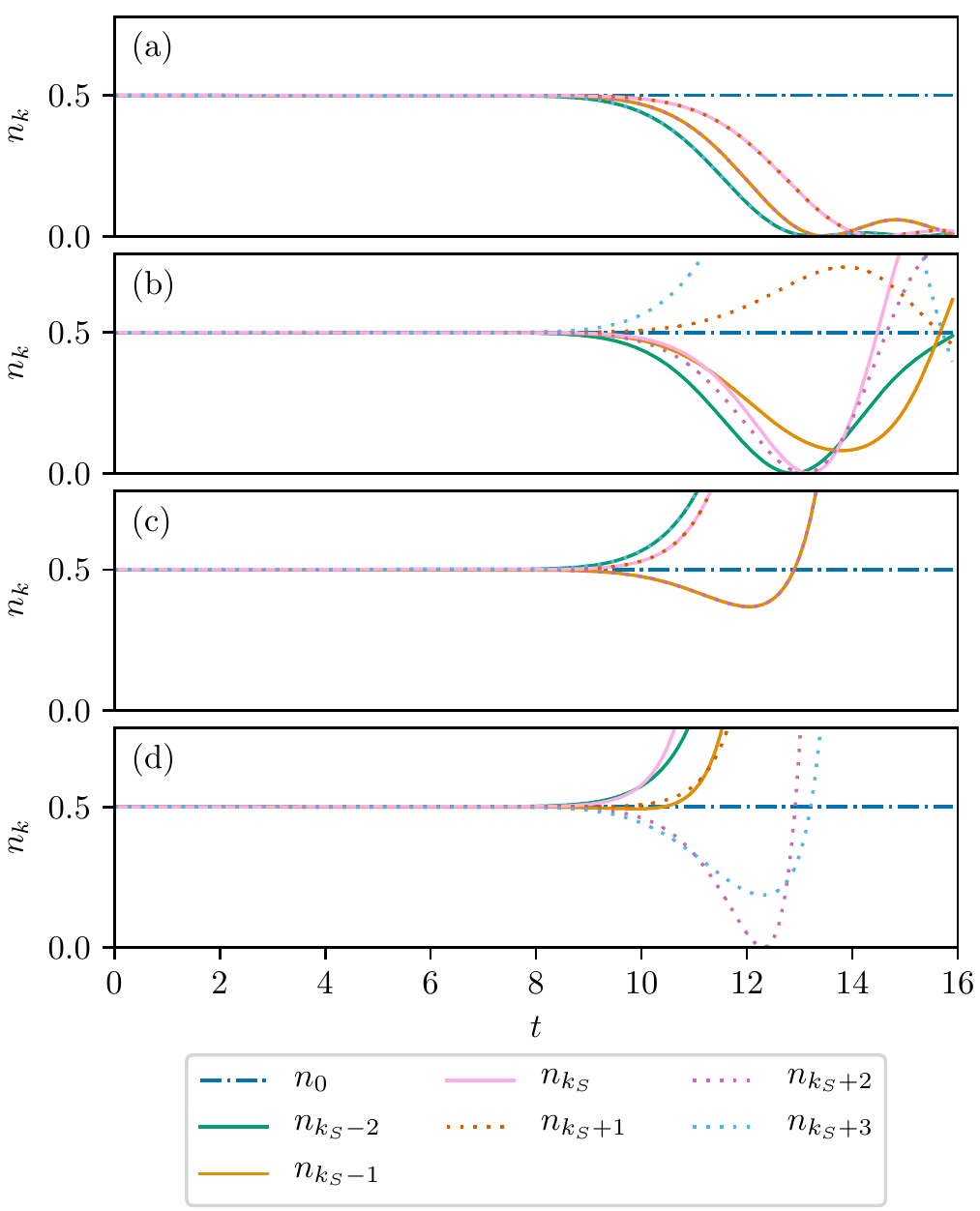}
  \caption{The particle numbers $n_k(t)$ of the inner 6 lattice sites for $\gamma=1$ corresponding to the four possibilities of the currents $\jR$ and $\jL$ in \cref{tab:current_possibilities_jL_j_jR}. The nonlinearity is set to zero, $g = 0$.}
  \label{fig:fig3}
\end{figure}

A comparison of the particle numbers $n_k(t)$ of the inner 6 lattice sites with $\gamma=1$ is illustrated in \cref{fig:fig3} for all four possibilities of the currents $\jR$ and $\jL$ in \cref{tab:current_possibilities_jL_j_jR}. The local maxima in \csubref{fig:fig3}{a} do not appear here because the coupling constant $\gamma = 1$ makes the tunneling current and the initial current equally large. For all four cases in \cref{fig:fig3} the occupations of the subsystem deviate from the constant value $n_0$ at $t\approx 8$, which is followed by chaotic dynamics. The second case in \csubref{fig:fig3}{b} represents the trivial case with identical initial phases at all lattice sites and vanishing loss and gain terms because the currents of the reservoir sites effectively supply a balanced gain and loss term for the subsystem. In \csubrefs{fig:fig3}{c} and \labelcsubref{fig:fig3}{d} the particle numbers of the inner 6 lattice sites diverge after the quasi-stationarity breaks down at $t \approx 8$. As in \cref{eq:deriv_n_k_general} the gain term $\gamma_{\kS}<0$ increases the occupation $n_{\kS}$ exponentially. The gain and loss terms $\gamma_{\kS} = - \gamma_{\kS+1}$ in the fourth case look similar to those of the \ac{TMS} in \cref{eq:GPE_2}, but due to the interaction $J\ne 0$ with the adjoining sites, the dynamics of the subsystem finally collapses due to the finite reservoir. The time where the quasi-stationarity breaks down due to the emptying of the outer wells increases with the number of wells used as reservoirs. This time scale decreases for nonlinear interactions $g \neq 0$ as our investigations showed. However, since the nonlinearity does not lead to qualitative differences with respect to the realizability or the dynamics, we will focus on the linear case.

Due to the difficult realization of a gain term with $\gamma_{\kS}<0$ in an actual experiment (e.g.\ see \cite{Robins2008}), we consider the first two cases in \cref{tab:current_possibilities_jL_j_jR} with the respective dynamics in \csubrefs{fig:fig3}{a} and \labelcsubref{fig:fig3}{b} as experimentally accessible situations with {\PT}-symmetric characteristics.

\subsubsection{Discussion of the phases}

\begin{figure}
  \centering
  \includegraphics[width=\columnwidth]{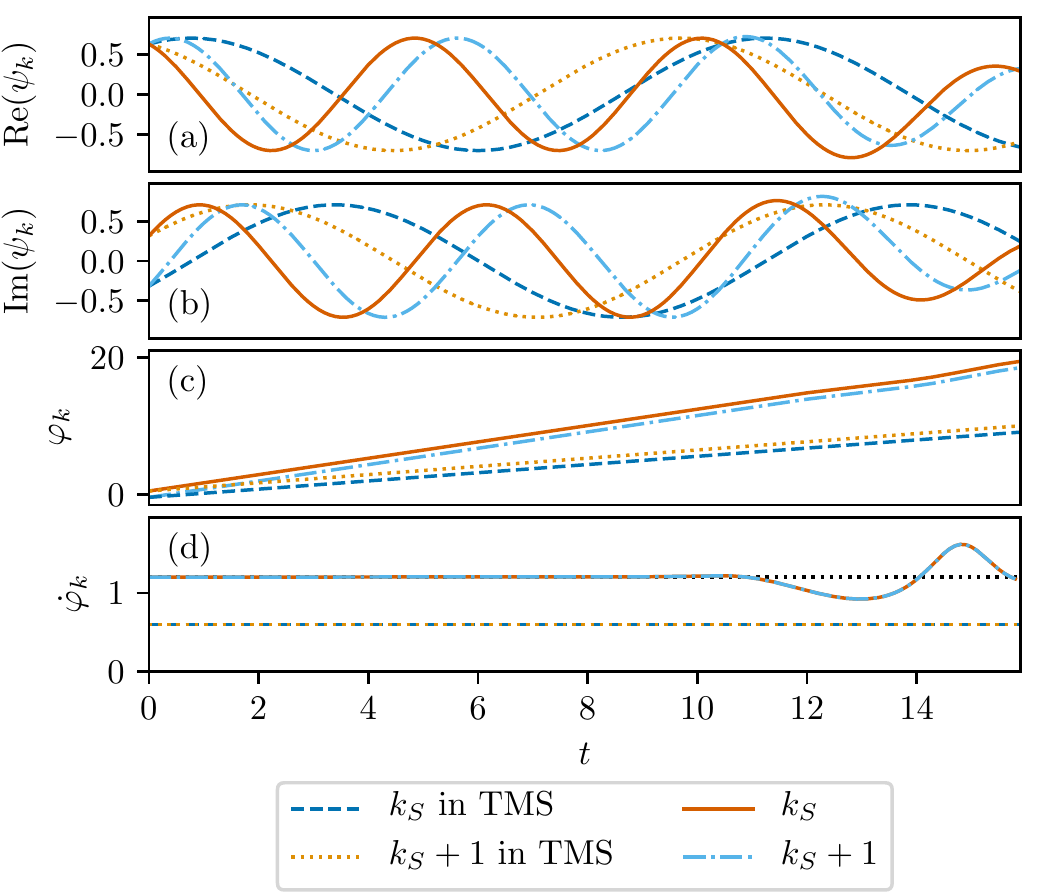}
  \caption{The real and imaginary parts of the wave functions of the subsystem, as well as their phases and the derivatives of these phases are compared with the expected values of the \ac{TMS}. The parameters are the same as in \cref{fig:fig2}.}
  \label{fig:fig4}
\end{figure}

Although the physical observables match the theoretically expected values, we will now consider the wave functions of the subsystem directly. To observe the oscillations of the phases it has to be ensured that the chemical potential does not vanish as for $\gamma = 1$. Therefore, in the following we use $\gamma = 0.8$.

The real and imaginary parts of these wave functions shown in \csubrefs{fig:fig4}{a} and \labelcsubref{fig:fig4}{b} are qualitatively similar. They start at the correct values but evolve differently in time. Therefore, since the norms $\abs{\psi_k(t)}^2=n_k(t)$ coincide with those of the \ac{TMS}, the time-dependent phases $\varphi_k(t)$ of the wave functions $\psi_k(t) = \sqrt{n_k(t)} \exp(\imag \varphi_k(t))$, which are shown in \csubref{fig:fig4}{c}, have to be analyzed. While the phase differences of the subsystem and \ac{TMS} are identical, their time derivatives $\dot{\varphi}_k(t)$ shown in \csubref{fig:fig4}{d} are different.

Since the wave functions of the \ac{TMS} evolve in time with the factor $\exp(-\imag \mu t)$, the derivatives of their phases $\dot{\varphi}_i(t) = -\mu = 0.6$ are half as large as those of the subsystem with the value $\dot{\varphi}_k(t<8) = 1.2$. In \cref{App:energyeigenvalue} it is shown that the chemical potentials indeed differ from one another by a factor of 2.

For the general case of a non-vanishing nonlinearity $g \neq 0$ the wave functions of the subsystem have the form $\psi_k(t) = \psi_k(0)\exp(-\imag \tilde{\mu}t)$ with the chemical potentials
\begin{align} \label{eq:tilde_mu}
	\tilde{\mu} = gn_0 \pm 2 \sqrt{J^2-\gamma^2},
\end{align}
whereas the wave functions of the \ac{TMS} evolve in time with the chemical potentials \labelcref{eq:mu_theo}. Since the currents and correlations in \cref{eq:def:c_kl,eq:def:j_kl} only depend on the phase differences, and the particle numbers are independent of any phases, the wave functions exhibit the same physical dynamics as those of the \ac{TMS}. Due to no additional degrees of freedom in \cref{subsubsec:PT-symm-regime}, the chemical potentials $\tilde{\mu}$ of the subsystem are not adjustable.

\subsection{Realizing broken {\PT} symmetry}

\begin{figure}
  \centering
  \includegraphics[width=\columnwidth]{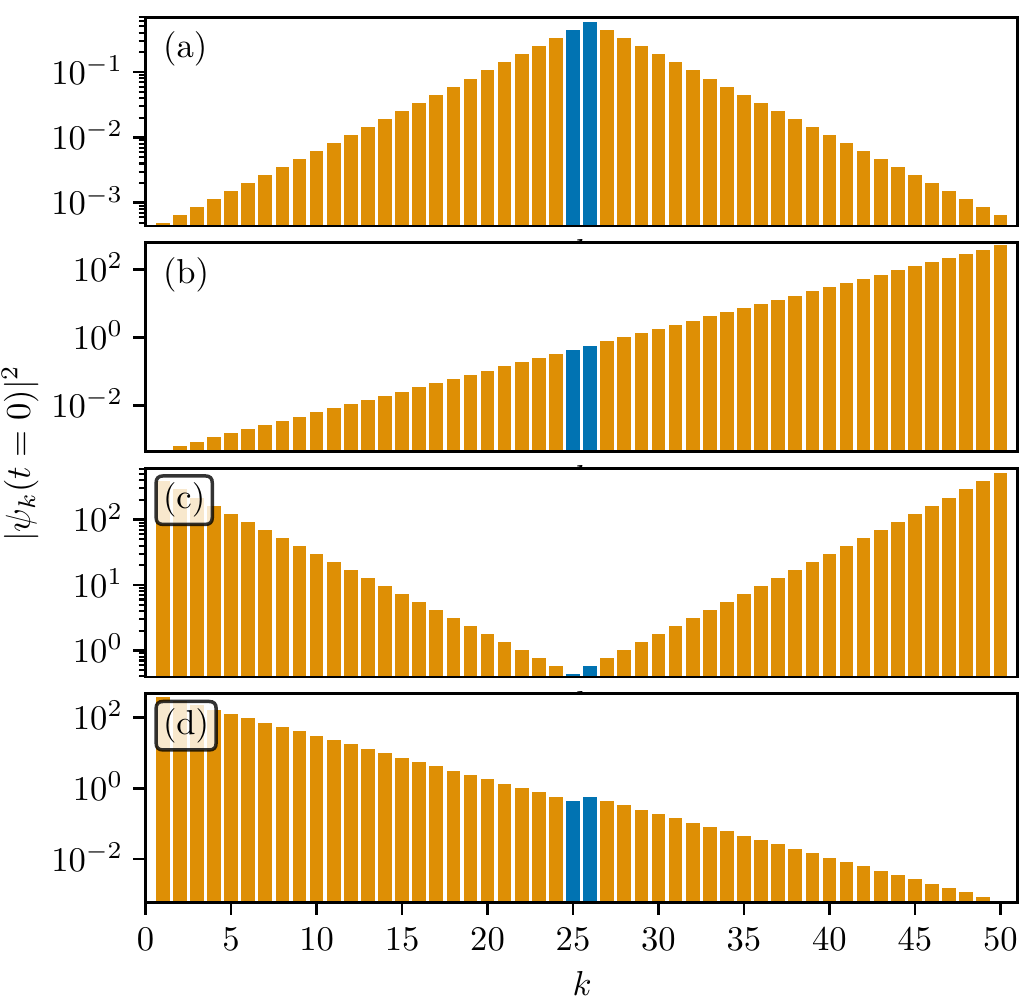}
  \caption{The initial occupations $\abs{\psi_k(t=0)}^2$ of the four possible decaying states with $\mu < 0$ for the values of the currents from \cref{tab:current_possibilities_jL_j_jR} with $\gamma=1.01$. The nonlinearity is set to zero, $g = 0$.}
  \label{fig:fig5}
\end{figure}

The aim of this subsection is to realize the {\PT}-symmetry broken states of the \ac{TMS} using the same setup as in the previous section. As derived in \cref{subsubsec:PT-broken-regime}, the approach of complex phases leads to exponentially distributed initial occupations (see \cref{eq:exp-distr-particle-num}), which are illustrated in \cref{fig:fig5} for the four cases in \cref{tab:current_possibilities_jL_j_jR} with $\gamma = 1.01 > J$. All four initial particle numbers of the subsystem are set to the same values given by \cref{eq:ansatz1_n_ks_n_ksp1} with $n_0 = 0.5$, where the sign belonging to the exponentially decaying occupations with $\mu<0$ is chosen. The nonlinearity is set to $g=0$ again, as a non-zero value disturbs the desired dynamics in \cref{eq:open_few-mode_model} because of the time-dependent norm $\abs{\psi_k}^2$. We want to emphasize that, although $n_{\kS}$ and $n_{\kS+1}$ are the same in every case shown in \cref{fig:fig5}, the case with just localized loss (cf.\ \csubref{fig:fig5}{a}) requires only a surprisingly small overall number of particles in the condensate. Therefore, this approach seems particularly suitable for an experimental realization.

\begin{figure}
  \centering
  \includegraphics[width=\columnwidth]{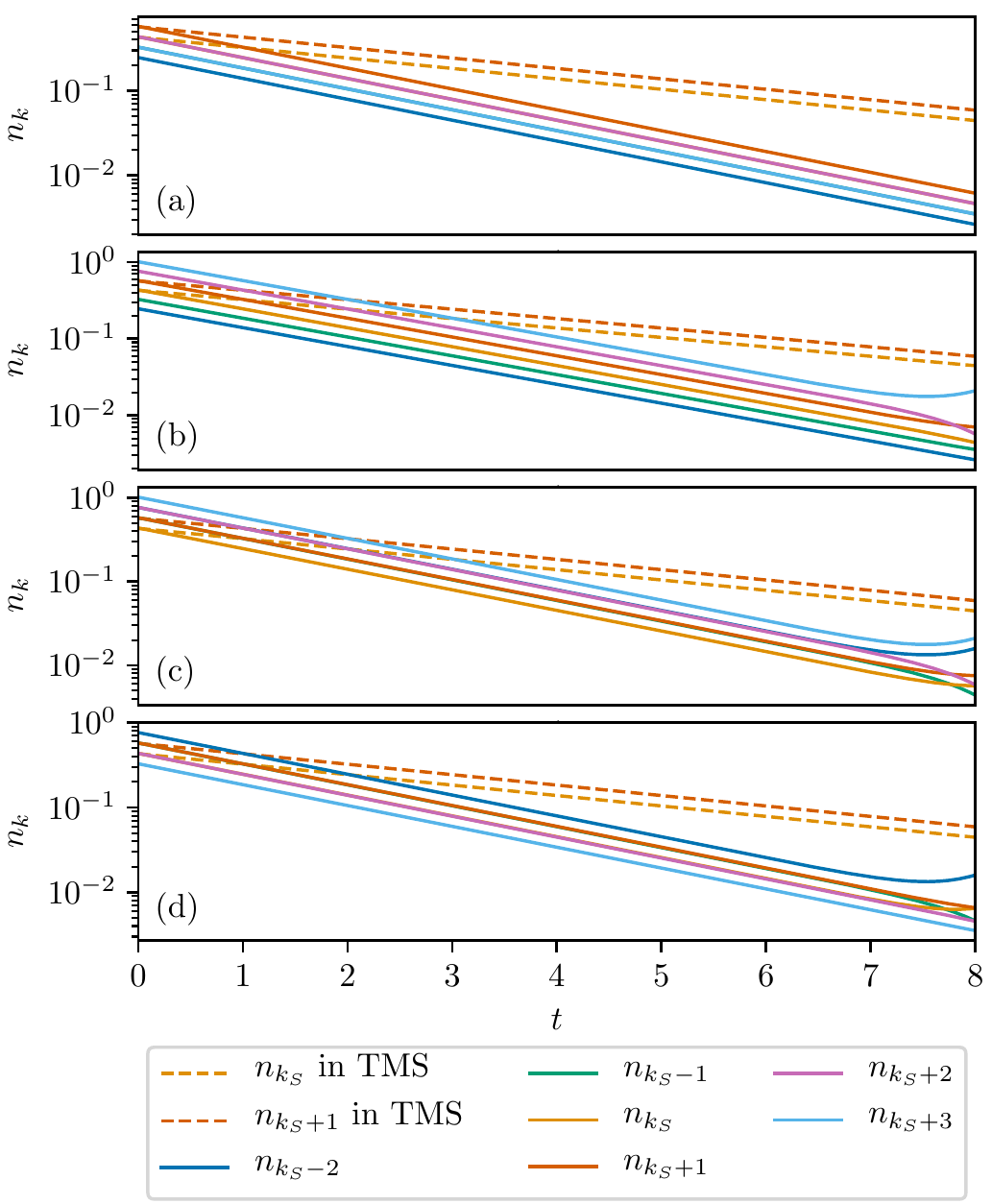}
  \caption{The particle numbers $n_k(t)$ of the inner 6 lattice sites for $\gamma=1.01$ correspond to the four possible decaying states with the initial occupations shown in \cref{fig:fig5}. The dash-dotted lines mark the expected particle numbers of the \ac{TMS}.}
  \label{fig:fig6}
\end{figure}

A comparison of the particle numbers $n_k(t)$ of the inner 6 lattice sites with $\gamma=1.01$ is illustrated in \cref{fig:fig6} for all four possibilities of the currents $\jR$ and $\jL$ in \cref{tab:current_possibilities_jL_j_jR}. It is noticeable that the occupations of the inner lattice sites evolve purely exponentially in time, but with a different factor compared to the \ac{TMS} in \cref{eq:n_i(t)_theo-broken}. This stable behavior collapses roughly at the same time $t \approx 8$ as the {\PT}-symmetric states in \cref{fig:fig3}.

A closer examination reveals that the exponentially decaying rates of these curves are identical in all subplots, and also twice as large as the ones of the \ac{TMS}. As a consequence, the wave functions exhibit the chemical potential $\tilde{\mu}$ as the {\PT}-symmetric wave functions in \cref{eq:tilde_mu} with $g = 0$. This complex eigenvalue results in time-dependent occupations
\begin{align}
	n_k(t) = n_k(0) \mathrm{e}^{2 \Im(\tilde{\mu})t},
\end{align}
and is thus consistent with the wave functions of the {\PT}-symmetric regime.

\section{Conclusion}

In this paper we proposed an experimental setup for the realization of a two-mode quantum system consisting of Bose-Einstein condensates coupled to a reservoir, which leads to localized particle gain and loss. With a suitable choice of the particle numbers and phases, respectively, in each well the system exhibits either {\PT} symmetry or broken {\PT} symmetry for a finite time interval. In contrast to previous work, the theoretical treatment of the experiment is based on an open few-mode model with time-independent uniform optical characteristics.

All suitable initial phases allow for four types of environments, some with additional gain and loss in the inner wells, in which quasi-stationary states can be produced in a subsystem having almost the same dynamics as the {\PT}-symmetric \ac{TMS}. However, a closer inspection shows that the chemical potentials of both systems in the linear case are proportional by a factor of two. Thus, real and imaginary parts of the wave functions in the inner wells oscillate with different frequency. By continuing the phases of the {\PT}-symmetric wave functions into the complex plane we find states with a time-dependent norm showing similar characteristics as the {\PT}-symmetry broken solutions of the \ac{TMS}. The resulting initial exponentially distributed particle numbers and equal absolute phase differences lead to wave functions with complex eigenvalues causing an exponential increase or decrease of the particle numbers in the inner wells, which differ from the behavior of the \ac{TMS} again by the same factor of two.

For an actual experimental realization the situation with only localized loss seems particularly suited, as this approach can significantly reduce the number of particles required. Such localized losses can simply be created via a focused electron beam as shown in Ref.~\cite{Labouvie2016}. The creation of arbitrary occupations in each lattice site is also experimentally possible \cite{Peil2003, Wuertz2009}. However, the preparation of specific phase differences between Bose-Einstein condensates in neighboring sites, which is crucial for our approach, remains demanding. A possible experimental technique for such phase engineering may be to optically imprint the phases via far-off resonant lasers \cite{Dobrek1990, Denschlag2000, Martellucci2007, Song2012}.



\appendix
\section{Comparison of the chemical potential of the TMS and the subsystem in the lattice}
\label{App:energyeigenvalue}

Here we show that the chemical potential of the subsystem in the lattice is twice as large as the one of the \ac{TMS} assuming that the nonlinearity vanishes, $g=0$. The wave functions of the lattice site $k$
\begin{align}
  \psi_k = \psiR + \imag \psiI ,
\end{align}
leaving out the indices due to clarity, is split into its real and imaginary parts. Thus the time derivative of the phase $\varphi=\arctan(\psiI / \psiR)$ can be calculated,
\begin{align}
  \dot{\varphi} = \frac{1}{1+(\psiI/\psiR)^2} \frac{\dot{\psiI}\psiR
    - \psiI\dot{\psiR}}{\psiR^2}
  =\frac{1}{n} \left( \dot{\psiI}\psiR - \psiI\dot{\psiR} \right) ,
  \label{eq:appendix_dot(varphi)}
\end{align}
with the particle number $\psiR^2 + \psiI^2 = n$. By using \cref{eq:GPE_2,eq:varphi_theo}, the derivatives of the phases of the \ac{TMS} yield
\begin{align}
  \dot{\varphi_1} = \dot{\varphi_2}
  &= J \cos(\varphi_2 - \varphi_1)
  = J \cos( \arcsin(\frac{\gamma}{J}) ) \notag\\
  &= \sqrt{J^2-\gamma^2} = -\mu ,
\end{align}
which are equivalent to the expected value in \cref{eq:mu_theo} for $g=0$. In the same manner the derivatives of the phases at the initial time $t = 0$, where all phase differences between neighboring sites have the same absolute value, can be calculated for a lattice described by
\begin{align}
  \imag \pdv{t} \psi_k
  =&	-J\psi_{k-1} - J\psi_{k+1} - \imag \frac{\gamma_k}{2}\psi_k .
\end{align}
Using the loss term $\gamma_{k_S+1} = 4\gamma$, for example, one finds
\begin{align}
  \dot{\varphi}_{\kS} = \dot{\varphi}_{\kS+1} = 2 J \cos( \varphi_{\kS+1} - \varphi_{\kS}) = -2\mu = -\tilde{\mu},
\end{align}
which explains the different behavior of the phases in the lattice system and in the \ac{TMS} shown in \csubrefs{fig:fig4}{c} and \labelcsubref{fig:fig4}{d}.


%

\end{document}